# Measuring Cognitive Conflict in Virtual Reality with Feedback-Related Negativity

Avinash Kumar Singh, Hsiang-Ting Chen, Jung-Tai King, Chin-Teng Lin
University of Technology Sydney

## ABSTRACT

As virtual reality (VR) emerges as a mainstream platform, designers have started to experiment new interaction techniques to enhance the user experience. This is a challenging task because designers not only strive to provide designs with good performance, but also carefully ensure not to disrupt users' immersive experience. There is a dire need for a new evaluation tool that extends beyond traditional quantitative measurements to assist designers in the design process.

We propose an EEG-based experiment framework that evaluates interaction techniques in VR by measuring *intentionally elicited* cognitive conflict. Through the analysis of the feedback-related negativity (FRN) as well as other quantitative measurements, this framework allows designers to evaluate the effect of the variables of interest.

We studied the framework by applying it to the fundamental task of 3D object selection using direct 3D input, i.e. tracked hand in VR. The cognitive conflict is intentionally elicited by manipulating the selection radius of the target object. Our first behavior experiment validated the framework in line with the findings of conflict-induced behavior adjustments similar to those reported in other classical psychology experiment paradigms. Our second EEG-based experiment examins the effect of the appearance of virtual hands. We found that the amplitude of FRN correlates with the level of realism of the virtual hands, which concurs with the Uncanny Valley theory.

**ACM Classification Keywords**
H.5.2. User Interfaces: Evaluation/methodology

## INTRODUCTION

Recent advances in display and tracking technologies bring affordable and plausible VR experience to the mass market. Despite the long history of VR, this is the first time we see a large quantity of creative and interactive content has been designed and produced specifically for the VR from scratch; being able to experience the content in VR is no longer a bonus feature or afterthought.

This paradigm shift from traditional passive visual stimulus to a more active immersive experience requires designers to reinvent new symbols and languages to facilitate communication and interaction in VR. This transition is challenging because these new interactions not only have to achieve good performance but should also avoid disrupting users' immersive experience.

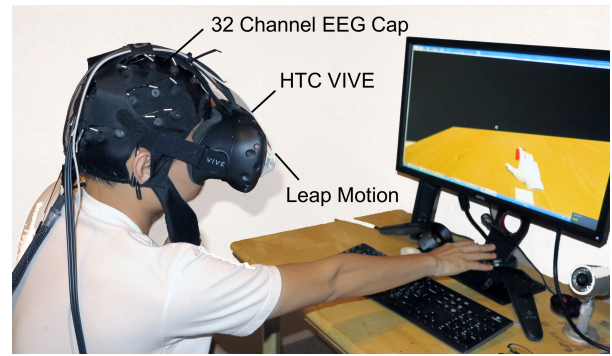

**Figure 1. Our EEG-based experiment framework evaluates the interaction techniques in VR by measuring intentionally elicited cognitive conflict.**

A range of measurements and visualization tools assists designers in the evaluation of the objective characteristics of an interaction. However, for the subjective measurements that are important in many VR scenarios, such as level of presence, focus, or emotions etc., designers still rely mainly on questionnaires and interviews, which can only be conducted at a particular time and cannot reliably address the changing dynamics of the interaction [26].

In searching for a continuous and reliable measurement of subjective characteristics, researchers have suggested leveraging brain imagining techniques, such as electroencephalography (EEG) and functional near-infrared spectroscopy (fNIRS), to inform the interaction design [19, 48]. As off-the-shelf EEG and fNIRS headsets are increasingly made available to the mass market by companies like Emotiv [14], Mindo [41], and Artinis [3], we believe brain imaging will become a convenient and essential tool for interaction design.

## COGNITIVE CONFLICT IN VR

This paper proposes an EEG-based experiment framework that evaluates the interaction techniques in VR by measuring *intentionally elicited* cognitive conflict. Previous works in the HCI community have mainly measured cognitive conflict in the background and used this additional data stream to recognize potential interaction errors [19, 54]. In contrast, our framework actively induces cognitive conflict by adjusting the variables of interest. Through the analysis of the feedback-related negativity (FRN) and other quantitative measurements, we correlate the amplitude of the conflict to the variables, which serve as a subjective indicator to the variables of interest.

We demonstrate the framework by the fundamental task of 3D object selection with direct 3D input, i.e. tracked hands (Figure 1). The cognitive conflict is intentionally elicited by manipulating the *selection radius* of the target object (Figure 2). For example, in the condition where the *selection radius* is larger than the actual radius, the object will go into the selection state, e.g. change its color, before the user's hand reaches it. In a realistic scenario, this kind of conflict occurs when the tracking precision is low or users experience poor depth perception as a result of inappropriate viewpoints or uncalibrated stereo displays.

We conducted two experiments to study and verify the proposed framework. The aim of the first behavior experiment was to identify behavior change induced by the conflict. The results concur with the behavior changes found in previous classical psychology experiments such as the Stroop test [34]: the conflict affects the reaction time and there is a first-order congruent effect. The second experiment applied the proposed framework to examine the effect of the appearance of virtual hands. The results show that the amplitude of FRN correlates with the level of realism of the virtual hands, complying with the Uncanny Valley theory [42], which states that the more realistic the virtual body is, the more sensitive users will be to imperfections or errors related to the interaction.

The main contributions of this paper are as follows:

- We propose a novel EEG-based experiment framework that assists the interaction design in VR by intentionally eliciting cognitive conflict and present a concrete experiment design for 3D object selection in VR.
- We show that the conflict-induced behavior changes for 3D object selection in VR are similar to those found in classical psychology experiment paradigms such as the Stroop test.
- We apply the proposed experiment framework to examine the effect of the appearance of virtual hands and find a correlation between the amplitude of conflict and the level of realism of the hand's appearance.

**RELATED WORK**

The nature and source of conflict varies, but generally speaking, conflict arises when competing or incompatible options are presented to an individual. Researchers in different communities study conflict from different perspectives such as psychology [9], cognitive science [13] and neuroscience [36].

This related work section focuses on the previous research works that are most relevant to our research questions and research methodology. We categorize these works into three types: *conflict due to competing options*, *conflict due to visual appearance*, and *conflict due to unexpected error*, and include a sub-section on the use of brain-computer interface in the HCI community.

**Conflict due to Competing Options**

Psychologists have constructed and studied a variety of conflict paradigms. Among them, the Stroop test [34] is perhaps the most well-studied. In this test, participants were instructed to respond to the names of colors printed in different color inks. The incongruent (conflict) condition arises when the name of the color differs from the color of its printed ink, while the congruent (no conflict) condition occurs when the name of the color is printed in the same color ink. In the incongruent case, the processing pathways for reading out words and naming the color of the ink compete with each other and result in a conflict. In the Eriksen flanker test [15], conflict arises when the stimulus is surrounded by two different spatial responses (left or right arrows). Another prominent paradigm, the Simon test, generates conflict by creating a spatial mismatch between a stimulus and the required response.

Previous experiments have also identified that the level of conflict affects both current and sequential trials. The congruency sequence effect [13, 36] states that the effect is smaller following an incongruent trial than a congruent one. These results triggered our interest in investigating how conflict affects the performance of fundamental interactions such as object selection in VR, and thus the first experiment.

**Conflicts due to Visual Appearance**

Computer graphics researchers have looked into the complex interplay between a rendered image, animation, and human perception. For example, McDonnel et al. [37] investigated the impact of rendering style on viewers' perception of virtual humans, and Hodgins et al. [24] studied how the degradation of human motion affects viewers' emotional responses. Most findings support the Uncanny Valley theory [42], which suggests that human-like robots are agreeable until they approach, but fail to attain, a lifelike appearance, at which point humans feel strong unease or possibly revulsion at even small imperfections.

Psychologists and neuroscientists have also examined brain activity when the subject is presented with objects that have different levels of realism. Perani et al. [45] showed subjects videos of both real and virtual hands in VR and 2D TV. The Positron Emission Tomography (PET) result showed that only real actions in the natural environment activate a visuospatial network that includes the right posterior parietal cortex. Han et al. [27] reported the functional Magnetic Resonance Imaging (fMRI) result of a higher level of brain activity when subjects watched a live action movie compared to a cartoon movie. More recently, Saygin et al. [46] utilized the fMRI repetition suppression methodology and identified a stronger effect when subjects watched the movement of a human-like robot compared to watching a human or a robot, which supports the Uncanny Valley theory.

We have so far described experiments in which participants passively watched pre-rendered videos throughout the sessions. Researchers have also studied participants' reactions in scenarios where the subjects actively perform tasks in the virtual environment with altered virtual appearance. Yuan and Steed [58] measured the galvanic skin response (GSR) and reproduced the rubber hand illusion [8] in an immersive virtual reality. Interestingly, they also reported that the illusion can be negated by replacing the virtual hand with an abstract cursor. Lin et al [32] investigated six distinct hand styles in the rubber hand illusion experiment and compared the induced threat levels. The questionnaire feedback suggests that subjects experience similar levels of virtual body ownership regardless of visual appearance.

**Conflicts due to Error**
Conflict also arises when the user is aware of making an error or the application does not behave as expected. In the context of 2D object selection, the HCI community has accumulated profound knowledge about the prediction and modeling of errors [57] as well as how users adjust their behavior to balance the tradeoff of error and time [5, 21].

Beyond standard measurements such as time to complete task, accuracy and error rate, Vi et al. [54, 55] proposed using the brain potential of Error Related Negativity (ERN) recorded from a consumer-level EEG headset to detect the occurrence of one's own or others' errors. Our work also leverages similar conflict due to unexpected feedback (FRN), but our framework intentionally induces the conflict and is in the completely different context of 3D object selection in VR.

Feedback-related negativity (FRN) is a negative component of event-related potential that occurs 250 to 300ms after a negative feedback stimulus [39]. Similar to the ERN, the FRN has the same neural source at the anterior cingulate cortex which reflects the same reinforcement learning signal generated in response to unexpected negative outcomes and performance monitoring [22]. Therefore, this kind of index could be used to explore the conflict that violates human expectation (e.g. the discrepancy between visual perception and motion graphics) in VR environment

**Brian-Computer Interface in HCI**
In the last decade, HCI researchers have explored the use of non-invasive brain imaging technologies, such as EEG and fNIRS, as interfaces to provide computer applications with the cognitive state of the user. Zander et al. [62] categorized the brain-computer interface (BCI) into three types: *active*, *reactive* and *passive*. *Active* BCI directly maps a user's brain pattern to a specific input command, e.g. control the mouse cursor with thought [16]. *Reactive* BCI leverages the brain's response to external audio or visual stimulus as input to the system, e.g. P300 speller [12]. The *Passive* BCI system translates brain activity without voluntary control into high-level cognition feedback, such as emotion, attention, etc., usually for the purpose of monitoring or evaluation [18, 19].

Most works from the HCI community utilize *reactive* and *passive* BCI, as rightfully predicted by Tan and Nijholt [10]. A series of works [1, 2, 23, 44, 48, 49, 50, 61] were recently undertaken, based on a relatively new brain imaging technology of functional Near-Infrared Spectroscopy (fNIRS). Solovey et al. proposed a guideline for using fNIRS in the HCI setting [48] and sequentially built and evaluated an interactive human-robot system [49, 50] which is assisted by a continuous supplement input-stream translated from the user's brain activity pattern. Afergan et al. [1, 2] used fNIRS to measure the workload as well as improve the target selection performance. Peck et al. [44] demonstrated that fNIRS is a valid tool for measuring the impact of visual design. Lastly, the BACh system [61] adaptively adjusts the difficulty level of music learning tasks based on the cognitive state of the learner. We refer readers to a more complete survey article by Yuksel et al. [60], due to limited space.

Before the emergence of fNIRS, electroencephalogram (EEG) had already been widely adopted by the HCI community [18]. The characteristic of high temporal resolution makes EEG particularly suitable for interaction that involves an immediate user response. Lee et al. [30] leveraged mismatch negativity and P3a to evaluate audio notification in realistic environments with ambient sounds, when users faced different levels of workload. Cherng et al. [11] studied user perceptions of graphics icons and provided an EEG-based evaluation of the semantic distance between icons. Yuksel et al. [59] built a P300-based BCI based on a multi-touch surface. The surface generates different flash patterns and elicits event related potentials in the EEG signals which can be used for various tasks such as object selection.

**RESEARCH GOALS AND EXPERIMENTS DESIGN**
The research goal of this paper is to investigate the feasibility of leveraging the measurement of cognitive conflict to assist the design of interaction in VR. We assume that if the interaction technique elicits less cognitive conflict or generates fewer unexpected responses, the less likely it is that it will disrupt users' sense of being in the virtual environment. Using 3D object selection as an example, we approach this research goal with two experiments addressing two different perspectives:

First, we examine the possibility of reproducing the results of previous behavior studies in the Psychology community. As described in the related work section, there are several established and well-studied experiment paradigms in the Psychology community [13, 36]. These paradigms are elegant yet specific. A successful reproduction of behavior results in the context of 3D object selection in VR implies that future research might be able to draw from the results, e.g. computational model, from the vast number of studies conducted by psychologists and neuroscientists.

Second, we apply the proposed EEG-based experiment framework to investigate the effect of the appearance of virtual hands on the 3D object selection tasks. Researchers have long shown interest in the effect of the appearance style on the virtual embodiment. By measuring cognitive conflict with EEG, we hope to provide a more objective evaluation and create a link to previous research on hand style and virtual embodiment [32, 58].

In the following section, we first describe the apparatus used followed by the details of the two experiments.

## APPARATUS AND ENVIORNMENT SETUP

### VR Setup

Both experiments used HTC Vive [9] as the head-mounted display. Vive uses an OLED display with a resolution of 2160 x 1200 and a refresh rate of 90 Hz. The user's head position is principally tracked with the embedded IMUs while the external Lighthouse tracking system clears the common tracking drift with a 60 Hz update rate.

Participants' hand motions are tracked with a Leap Motion controller attached to the front of the HTC Vive. Leap Motion tracks the fingers, palms and arms of both hands up to around 60 cm above the device. The tracking accuracy is reported as 0.2mm [56] and the latency is around 30 milliseconds [29].

### EEG Recording and Processing

Participants were required to wear a wired EEG cap with 32 Ag/AgCl electrodes, including two reference electrodes (opposite lateral mastoids). The placement of the EEG electrodes was consistent with the modified international 10-20 system. The contact impedance between all electrodes and cortex was maintained below 5kΩ. The EEG recordings were collected using a Scan SynAmps2 Express system (Compumedics Ltd., VIC, Australia). The EEG recordings were digitally sampled at 1kHz with a 16-bit resolution.

An assistant helped participants put on the EEG cap first then the HMD. In the pilot study, we directly put the top belt of the HTC VIVE on top of the central channels of the EEG cap. Interestingly, since these EEG channels are pressed firmly onto the scalp, they actually provide cleaner signals. However, participants also found these firmly pressed EEG channels uncomfortable. Thus during the formal experiments, we manually adjusted the top belt of VIVE to avoid or reduce the pressure applied onto the EEG channels.

## EXPERIMENT 1: BEHAVIOR EXPERIMENT ON CONFLICT-INDUCED BEHAVIOR CHANGE

The first behavior experiment investigates whether it is possible to reproduce conflict-induced behavior changes in previous psychology experiment paradigms in the context of direct 3D input in VR. Based on previous behavior results [9, 13], we made two hypotheses:

- *H1*: time-to-complete should increase when conflicts happen, then slowly decrease as the user adapts to the conflicts.
- *H2*: the result should exhibit the first-order congruency effect, i.e. the reaction time of a conflict-trial preceded by another conflict-trial should be smaller than the reaction time of a conflict-trial preceded by a non-conflict-trial.

### Participants

Behavior data were collected from sixteen right-handed participants (male, age 20-26 years) recruited from the university. All participants have basic knowledge about brain computer interface and the form of behavior experiments. None had used head-mounted display nor leap motion before this experiment.

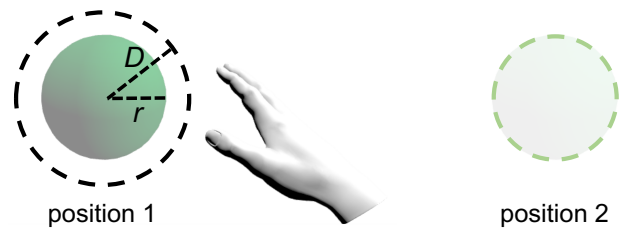

**Figure 2 the 3D object selection task of the first behavior experiment. The target object sequentially alternates between position 1 and position 2.**

### Experiment Design

Each participant performed a series of 3D object selection tasks with the direct input method, i.e. their tracked hand. In each trial, a semi-transparent green sphere appeared in front of the user (Figure 2). The sphere alternated sequentially between left and right positions in the trials. The sphere turned red when the participant's hand stayed within a distance $D$ of the center of the sphere and the target selection was completed when the hand stayed within $D$ for 0.3 seconds.

The experiment used a with-in subject design with $D$ as the sole independent variable with three levels: $D1 = 0.2r$, $D2 = r$, $D3 = 1.5r$, where $r$ is the actual radius of the sphere. Each participant performed 300 trials, 100 trials for each level. $D$ changed every 10 trials, which we defined as a *trial block*, in a random order. There was a 30 sec resting period after every three trial blocks. On average the experiment took about 10 to 15 minute.

### Results and Discussion

Figure 3 shows the average task completion time off all *trial block*. The graph concurs with *H1* and shows that the time to completion is much higher for the first two trials, right after the change of $D$, and gradually decrease toward the end of the block as the participant adapted to the changes. Interestingly, we also see the same phenomena in the condition without conflict, i.e. the selection radius is the same as the actual radius. It implies that the source of cognitive conflict for the first trial in the block also involves the change of the selection radius.

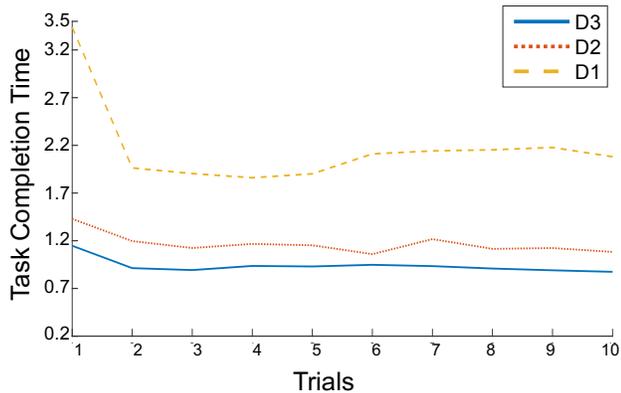

Figure 3. The average task completion time for the trial blocks with different D.

The first-order congruency effect states that the reaction time of an incongruent trial preceded by an incongruent trial should be smaller than an incongruent trial preceded by a congruent trial. In our experiment, only *D2* is congruent, where *D* matches *r*, while both *D1* and *D3* are incongruent. Figure 4 shows the trial completion time of the congruency combinations of interest, namely D2/D1 (congruent /incongruent), D3/D1 (incongruent / incongruent), D2/D3 (congruent / incongruent) and D1/D3 (incongruent / incongruent). First two bars show the trial time for the first three trials of the *D1* block and last two bars show the time-to-complete for the first three trials of the *D3* block. The result shows that first three trials in *D1* block preceded by *D3* is significantly faster than that preceded by *D2*. Similar significant effect was found for the block with *D3* condition. The result shows the first-order congruency effect and match the previous findings in other psychology experiments [13]. Thus *H2* is also correct.

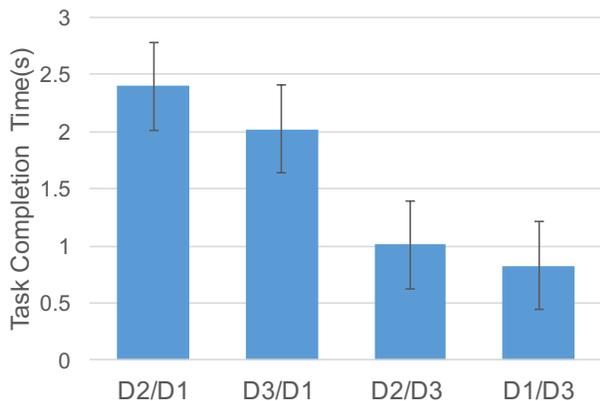

Figure 4 congruent / incongruent sequence for first three trials in the block

Note that we took the average time of the first three trials in the block instead of only the first trial because the main source of conflict for the first trial might be the change of radius instead of the mismatch between *D* and *r*.

In sum, experiment 1 suggests that although it is different in form, our experiment paradigm for 3D object selection reproduces the important findings of classical psychology experiment such as the Stroop test and Eriksen flanker test. It suggests that previous computational model in neuroscience experiments might also be applicable to our experiment framework.

**EXPERIMENT 2: EEG-BASED EXPERIMENT**
The second experiment applied the proposed EEG-based framework to evaluate the effect of the appearance of virtual hands to the task of 3D object selection in VR. We hypothesize that

- *H1*: The amplitude of FRN should correlate with hand style.
- *H2*: Participants would favor realistic-looking hand.

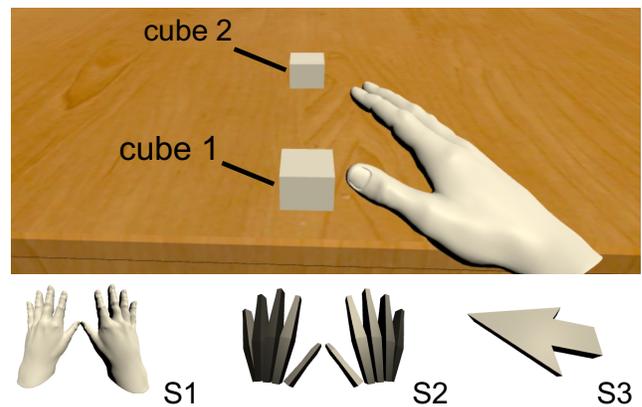

Figure 5. Top sub-figure shows the scene of the experiment 2. Each participant is instructed to put his/her hand at cube 1, then reaches for cube 2. Three sub-figures at bottom are the three hand styles used.

**Participants and Environment**
EEG data were recorded from 10 right-handed participants (male). The mean age was 22.7 years (in a range of 20-26 years) with no prior experience of the experiment. Following a detailed explanation of the experimental procedure, all subjects provided informed consent before participating in the study. This study had the institute's human research ethics committee approval and was conducted in a temperature controlled and soundproofed room. None of the participants had a history of psychological disorder which might affect the experiment results.

**Experiment Design**
Each participant performed the 3D object selection task with their tracked hands in VR. Figure 5 shows the experiment scenario. At the beginning of the trial, the participant would see two cubes on the table. The participant was instructed to first touch cube 1, then stretched her hand to the cube 2. The cube would turn red when it was selected. Participants need to finish each task within 5s and there was a 5s resting after each trial.

The experiment uses a 3 by 2 with-in subject design. Independent variables are the hand style (realistic hand, robotic hand, and 3D arrow, bottom row of Figure 5) and selection distance $D$ ($D1$, equals to the size of the cube and $D2$ is twice the size of the cube, similar to Figure 2). There are three 20-minute sessions with a 5-minute resting time in between. Each session corresponds to one hand style and consist of 120 trials. We used the oddball paradigm that shows three trials with condition $D1$ followed by a trial with condition $D2$. We counterbalanced the hand style among participants. On average the experiment took about two hours including the initial setup of EEG cap and HMD.

**Results**

ERP (event related potential) analysis has been done on the collected EEG data from the participants performing the object selection task. EEG data were filtered offline with 40 Hz low-pass and 1 Hz high-pass filters and further manually clear some clear artifacts. An epoch was defined from 200 ms prior to the stimulus and 500ms post stimulus.

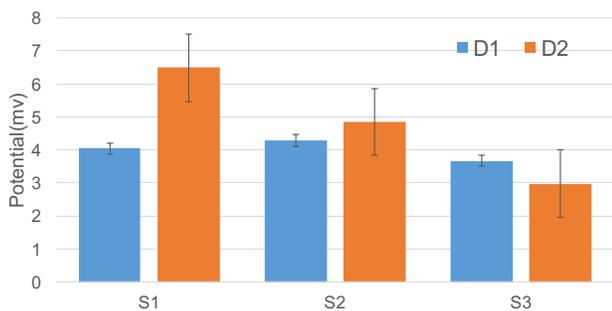

**Figure 6. ERP Area difference for 120ms-220ms. There are significant effects for S1 and S2 pairs.**

Extracted epochs are further analyzed to find the region of interest and average topographical map has been calculated. Topographical maps (Figure 8 bottom row) for each condition showed that front region of brain is the most significant ($p<0.05$) region of brain hence chosen for further analysis. Trials from the more frequent conditions (congruent condition) were randomly chosen to match the number of trials in the less frequent conditions (incongruent condition), thus controlling for differences in the signal to noise ratio (SNR) due to varying numbers of trials. ERP analysis has been performed to find the local minima or maxima over the electrode average in frontal region to find any event related negativity for all condition of trails. It was found that FRN (Figure 8 top row) of $D2$ has a significant difference against $D1$ for S1 and S2 condition whereas there was no significance for S3 condition.

Further analysis of event related activity has been done for a time range for 120ms-220ms to see if this event related negativity is because negative feedback due to conflict in participants. Figure 8 shows that participants showed the higher feedback related negativity (FRN) around 120-220ms during change in distance ($D2$) compare to normal distance ($D1$) for rendering of realistic hand style (S1) while FRN fall off more than half for rendering robotic hand (S2) for change in radius condition. One the other hand rendering of arrow hand style (S3) showed almost no FRN at all during change in distance.

The main purpose of the questionnaire is to understand how participants perceive the three different hand rendering style. Figure 7 showed both the questions and the results. All users consider hand style S1 is more realistic then two others ($p<0.05$). Surprisingly, the result shows that there is no significant difference between the suitability and preference between hand style S1 and S2. Some users actually suggested that prefer hand style S2 for the 3D object selection tasks. When asked about this, multiple participants suggested that the preference to $S2$ is mainly due to its occlude the target less. Also interestingly, for the last question about the level of conflict, participants ranked the level of conflicts as what FRN result shows (all with significant effect $p<0.05$).

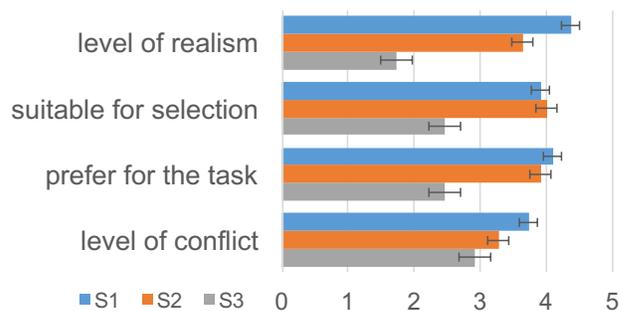

**Figure 7. questionnaire result for the EEG-based experiment.**

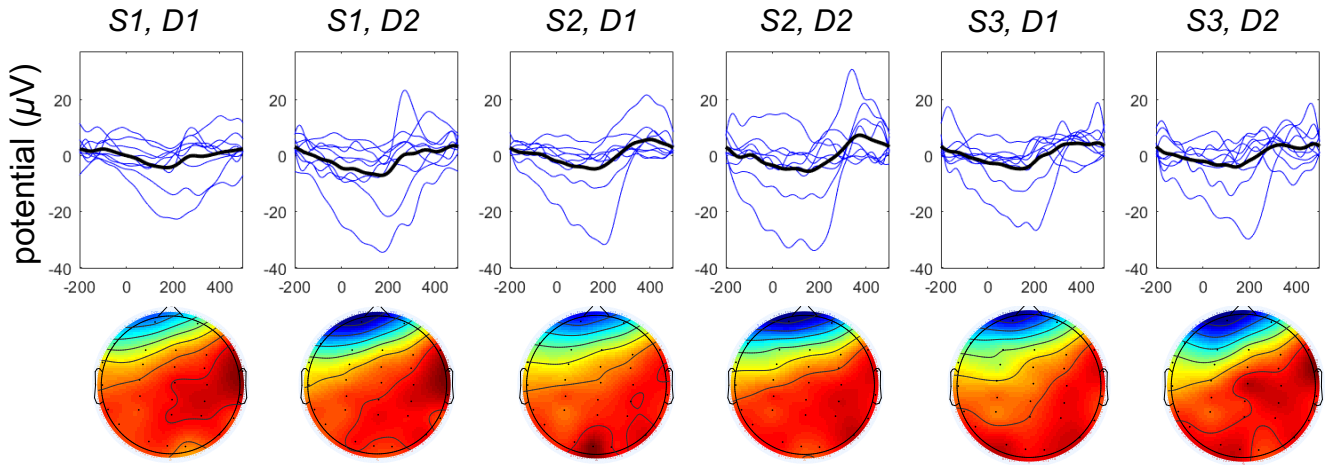

**Figure 8.** Top row: ERP from all participants. Black line is the average of all participants. Red line is the maxima FRN in the time window of 120 to 220 ms. Bottom Row: Topography shows frontal area as regions of interest.

**Discussion**

As hypothesized, both FRN and questionnaire suggests a correlation between the amplitude of FRN and the appearance of the virtual hand. This result echoes with the famous Uncanny Valley theory [42], which states that as a robot approaches, but fails to attain, the likable human-like appearance, there will be a point where users find even the slight imperfection unpleasant. In our case, as the virtual hand becoming more realistic looking, the participants also become more aware of the errors.

On the contrary, it is a bit surprising that there is almost no effect in FRN for the condition *S3*. The result implies that the participants are more tolerant or not very responsive to the error when they feel the virtual hand is less like a part of their body. The similar effect can also be found in the rubber hand illusion test [32, 51] where the participants felt less threatened to virtual threat, e.g. knife, saws etc., when their virtual body counterpart was not rendered realistically.

This finding implies that depends on the goals of the interaction and the hardware capability, higher rendering quality might not always be good. For example, if the tracking precision is likely to be compromised or the display quality of a HMD is not ideal, then using a less realistic rendering style might actually be helpful. On the contrary, if the nature of task and hardware permits, participants favor the more human-like looking of their virtual body.

**COGNITIVE CONFLICTS IN VR APPLICAITONS**

Previous sections show that the proposed experiment paradigm elicited conflict-induced behavior changes that were similar to previous psychology experiments and can therefore be used to evaluate the effect of the subject factor, i.e. the correlation between hand appearance and the amplitude of the conflict. The experiment paradigm can also be applied to evaluate other important factors relevant to interaction in VR.

**Evaluating the Importance of Different Factors for 3D Object Selection**

Researchers have long been curious about the relationship between levels of immersion and presence [7]. There have been many inspiring works in recent years that aim to add different sensory feedback into VR and interaction design [43]. For example, Impacto [33] renders the haptic feedback with both solenoid and electrical muscle stimulation, Level-Ups [47] adds a self-contained vertical actuator to the bottom of the foot, and HapticTurk [10] replaces the motion platform with humans. Most of these works rely on questionnaires and interviews to evaluate the effect of the feedback. However, most of them actually have a clear event, e.g. the time when haptic feedback or motion feedback is applied, and the amplitude of the cognitive conflict will be a useful tool for providing continuous user feedback to the system.

**Manipulating Sense**

The proposed experiment paradigm can also be used to evaluate the effectiveness and the range of recent works that manipulating senses to overcome physics constraints, such as limited number of props [4], limited space [52], and cyber sickness [17]. Again, in these cases, by controlling the source of conflicts, e.g. visual warping, we can estimate a reasonable range for subtle sense manipulation without getting noticed or causing discomfort.

**LIMITATION**

Our current experiment setup used the Scan SynAmps2 Express system and the recorded EEGs were analyzed off-line. Due to its long setup time, this device is only suitable for an initial investigation in a lab environment. We believe it should be possible to reproduce the result using off-the-shelf portable EEG devices [14, 41] and processing the data in real-time [20, 54].

During the experiment, we manually adjust the belt of HMD to avoid contact with the sensors on the EEG cap. It might not be possible if we are looking for the cap with higher sensor density. We believe the integration of the EEG cap and HMD is a natural one and we expect to see commercial products from companies like MindMaze [10] to be available to the market soon.

Synchronization is also a challenging issue for the hardware integration, especially if we are looking for specific signal such as N200 or P300. Leap Motion introduces a 30ms delay [29], both VIVE and Leap Motion has potential tracking precision error, the event generated from Unity 3D is limited by the rendering frame rate (60 FPS), and there is also another system delay for the communication between Unity and the parallel port of Scan (our EEG system). We estimate the latency is around 100 to 150 ms, which might also be the reason why we saw some negativity before event starts in Figure 8. For future works that focuses on specific ERPs such as N200 or P300, a dedicated synchronization hardware might be needed.

Finally, for well-defined tasks such as 3D object selection in VR, the appearance of cognitive conflict is mostly undesirable and might harm the sense of presence. However, for more complex tasks or interactive content, cognitive conflict might not always diminish the sense of presence. For example, cognitive conflict has long been used as a strategy to encourage students examining their previous knowledge and aiming for conceptual change [31]. We believe extending the framework to address such complex scenario is an interesting future research direction.

**CONCLUSION**

We proposed an EEG-based experiment paradigm that leverages the measurement of cognitive conflict to evaluate interaction techniques in VR. We demonstrated and validated the paradigm for the task of 3D object selection in VR using a direct 3D input method. The first behavior experiment shows that our paradigm induced similar behavior changes to previous psychology experiments in that paradigm. The second EEG-based experiment evaluated the effect of the appearance of a virtual hand and found that users become more sensitive to the conflict when the virtual hand is rendered in a realistic way, which concurs with the Uncanny Valley theory.